\documentclass[pra,twocolumn,shownopacs,amssymb]{revtex4}
\usepackage{graphicx}
\usepackage{amsmath}
\DeclareMathOperator{\cosech}{csch}
\DeclareMathOperator{\asinh}{arcsinh}
    
\begin{document}

\title{Bardeen-Cooper-Schrieffer--type pairing in a spin-$\frac{1}{2}$ Bose gas with spin-orbit coupling}

\author{M. Iskin}
\affiliation{Department of Physics, Ko\c{c} University, Rumelifeneri Yolu, 
34450 Sar\i yer, Istanbul, Turkey}

\date{\today}

\begin{abstract}

We apply the functional path-integral approach to analyze how the presence 
of a spin-orbit coupling (SOC) affects the basic properties of a BCS-type 
paired state in a two-component Bose gas. In addition to a mean-field theory 
that is based on the saddle-point approximation for the inter-component pairing, 
we derive a Ginzburg-Landau theory by including the Gaussian fluctuations 
on top, and use them to reveal the crucial roles played by the momentum-space 
structure of an arbitrary SOC field in the stability of the paired state at finite 
temperatures. For this purpose, we calculate the critical transition temperature 
for the formation of paired bosons, and that of the gapless quasiparticle 
excitations for a broad range of interaction and SOC strengths. In support of 
our results for the many-body problem, we also benchmark our numerical 
calculations against the analytically-tractable limits, and provide a full account 
of the two-body limit including its non-vanishing binding energy for arbitrarily 
weak interactions and the anisotropic effective mass tensor. 

\end{abstract}

\maketitle

\section{Introduction}
When atomic fermions transform into molecular bosons by way of many-body 
pairing, the mechanical stability of the paired state is inherently enforced by 
the Pauli exclusion principle, i.e., through a Hartree shift of the chemical potential
by inducing a pairwise interaction that is effectively weak and repulsive. 
This intrinsic stability is what lies behind the long-sought realization of the 
so-called BCS-BEC crossover, when a two-component Fermi gas is magnetically 
swept across a Feshbach resonance~\cite{ufg}. Having witnessed more than 
a decade of tremendous successes since their creation, the ultra-cold Fermi 
gases has become a thriving field in modern quantum physics as it keeps 
enriching its toolbox with a wide-range of applications for the strongly-correlated 
phenomena in a much broader context~\cite{giorgini08, strinati18}. 

In comparison, the analogous evolution from the BCS-type many-body paired 
state of atomic bosons to the BEC of molecular bosons is to a great extent 
an uncharted territory in a Bose gas. Despite many theoretical attempts dating 
back more than half a century~\cite{valatin58, evans69, evans73, nozieres82, 
jeon02, radzihovsky04, koetsier09, qiang10}, the crossover studies have 
been hindered by the natural tendency of paired state to a mechanical 
collapse in the lack of a bosonic counterpart for the exclusion principle. 
When a spinless Bose gas of atoms is magnetically swept across a Feshbach 
resonance, the lifetime of the resultant molecules turned out to be too short 
for reaching an equilibrium state with a molecular 
BEC~\cite{donley02, herbig03, xu03, thompson05}. 
To overcome this difficulty, it is quite clear that one needs to search for exotic 
Bose systems that may exhibit enhanced stability for the many-body pairing.
For instance, bosonic particles with an internal spin structure was introduced 
by Nozi\'{e}res and Saint James as an alternate~\cite{nozieres82}, and thoroughly 
analyzed for the spin-$1$ case.

Motivated by the recent creations of two-component quantum gases 
with two-dimensional SOCs~\cite{huang16, meng16, wu16, sun17}, here
we revisit this old-standing problem in a so-called spin-$\frac{1}{2}$ 
Bose gas. Having a dilute Bose gas with short-ranged density-density 
interactions in mind, we consider an inter-component attraction
$
U_{\uparrow \downarrow} = U_{\downarrow \uparrow} = - g < 0,
$
and analyze how the presence of a SOC affects the resultant pairing 
correlations~\cite{li14, luo17}. The mechanical collapse is counteracted 
by the Hartree terms arising from the intra-component 
repulsions~\cite{nozieres82, luo17, mueff, muB}. 
Then, assuming that the instability towards a BCS-like paired state is 
favored against the competing states, e.g., collapse, fragmentation, phase
separation, etc.~\cite{valatin58, evans69, evans73, nozieres82, jeon02, 
radzihovsky04, koetsier09, qiang10, li14, luo17}, one may treat the 
inter-component attraction through a close analogy with the theory of paired 
fermions~\cite{giorgini08, strinati18}. For this purpose, we apply the functional 
path-integral approach, and derive a mean-field theory that is based on the 
saddle-point approximation for pairing, and then a Ginzburg-Landau theory 
by including the Gaussian fluctuations on top. Our analysis suggests that 
while the SOC has a minor role in the strong-interaction limit where the 
ground state at zero temperature is a BEC of paired bosons, increasing 
its strength in the weak-interaction limit may allow for the creation of a paired 
state at much lower temperatures. We also provide a full account of the 
two-body problem including its non-vanishing binding energy for arbitrarily 
weak interactions and the anisotropic effective mass tensor.

\section{Three-dimensional spin-$\frac{1}{2}$ Bose gas with an arbitrary SOC}
We are interested in a two-component Bose gas that is described by the many-body 
Hamiltonian
\begin{align}
\label{eqn:ham}
H &= \sum_{ab\mathbf{k}} c_{a \mathbf{k}}^\dagger 
(\xi_\mathbf{k} \sigma_0 + \mathbf{S}_\mathbf{k} \cdot \boldsymbol{\sigma})_{ab}
c_{b \mathbf{k}} \\
&+ \frac{1}{2} \sum_{ab \mathbf{k} \mathbf{k'} \mathbf{q}} 
U_{ab} c_{a, \mathbf{k}+\mathbf{q}/2}^\dagger c_{b, -\mathbf{k}+\mathbf{q}/2}^\dagger
c_{b, -\mathbf{k'}+\mathbf{q}/2} c_{a, \mathbf{k'}+\mathbf{q}/2} \nonumber,
\end{align}
where the wavevector $\mathbf{k} = (k_x, k_y, k_z)$ labels the momentum 
eigenstates, and the pseudospin $a \in \{\uparrow, \downarrow\}$ labels the 
atomic components in such a way that $c_{a \mathbf{k}}^\dagger$ creates a 
pseudospin-$a$ boson with momentum $\mathbf{k}$ (in units of $\hbar \to 1$). 
Assuming the components are population-balanced, 
$
\xi_\mathbf{k} = \epsilon_\mathbf{k} - \mu
$ 
includes the parabolic dispersion $\epsilon_\mathbf{k} = k^2/(2m)$ of the 
particles in free space and their chemical potential $\mu < 0$~\cite{mueff, muB}. 
In addition, $\sigma_0$ is a $2\times2$ unit matrix,
$
\mathbf{S}_\mathbf{k} = (S_\mathbf{k}^x, S_\mathbf{k}^y, S_\mathbf{k}^z)
$
is a SOC field whose components $S_\mathbf{k}^i = \alpha_i k_i$ are controlled 
independently by the strengths $\alpha_i \ge 0$, and 
$\boldsymbol{\sigma} = (\sigma_x, \sigma_y, \sigma_z)$ is a vector of Pauli 
spin matrices. Below it is called an XYZ or a Weyl SOC if $\alpha_x = \alpha_y = \alpha_z = \alpha$ 
is isotropic in the entire $\mathbf{k}$ space, an XY or a Rashba SOC if 
$\alpha_x = \alpha_y = \alpha$ is isotropic in $k_xk_y$ plane with 
$\alpha_z = 0$, and a YZ SOC if $\alpha_y = \alpha_z = \alpha$ is isotropic 
in $k_yk_z$ plane with $\alpha_x = 0$.

\section{Mean-field theory for the inter-component pairing}
After a straightforward algebra, the saddle-point contribution $\Omega_0$ 
to the thermodynamic potential can be written as
$
\Omega_0 = A_0 + (T/2) \mathrm{Tr} \sum_k \ln [G_0^{-1}(k)/T].
$
Here,
$
A_0 = |\Delta_0|^2/g - \sum_{\mathbf{k}} \xi_\mathbf{k}
$
with the complex number $\Delta_0$ corresponding to the mean-field order 
parameter for the stationary pairs, and determined by the thermal average
$
\Delta_\mathbf{q} = -g \sum_\mathbf{k} \langle c_{\downarrow, -\mathbf{k}+\mathbf{q}/2} 
c_{\uparrow, \mathbf{k}+\mathbf{q}/2} \rangle
$
in the $\mathbf{q} \to \mathbf{0}$ limit. In addition, $T$ is the temperature with 
$k_B \to 1$ the Boltzmann constant, $\mathrm{Tr}$ is the trace, and $k$ denotes 
a combined summation index for $(\mathbf{k}, i\omega_\ell)$ 
where  $\omega_\ell = 2\pi T \ell$ is the bosonic Matsubara frequency for the 
particles with $\ell$ an integer.
Furthermore, 
\begin{align*}
G_0^{-1} = \left[
\begin{matrix}
(i\omega_\ell + \xi_\mathbf{k})\sigma_0 + \mathbf{S}_\mathbf{k} \cdot \boldsymbol{\sigma} & \Delta_0\sigma_x \\ 
\Delta_0^* \sigma_x & (-i\omega_\ell + \xi_\mathbf{k})\sigma_0 - \mathbf{S}_\mathbf{k} \cdot \boldsymbol{\sigma^*}
\end{matrix}
\right]
\end{align*}
corresponds to the inverse Green's function associated with the Nambu spinor
$
\psi_k^\dagger = (c_{\uparrow k}^\dagger, c_{\downarrow k}^\dagger, 
c_{\uparrow, -k}, c_{\downarrow, -k}).
$
A compact way to express
$
\Omega_0 =  A_0 + (T/2)\sum_{s s' k} \ln [(i\omega_\ell + s' E_{s \mathbf{k} })/T]
$
is through the quasiparticle energies
$
E_{s \mathbf{k} } = \sqrt{\xi_\mathbf{k}^2 - |\Delta_0|^2 + S_\mathbf{k}^2 
+ 2 s B_\mathbf{k}},
$
where $s \in \{+, - \}$,
$
S_\mathbf{k} = [(S_\mathbf{k}^\perp)^2 + (S_\mathbf{k}^z)^2]^{1/2}
$
is the strength of the SOC field with 
$
S_\mathbf{k}^\perp = [(S_\mathbf{k}^x)^2 + (S_\mathbf{k}^y)^2]^{1/2},
$
and 
$
B_\mathbf{k} = [\xi_\mathbf{k}^2S_\mathbf{k}^2 - |\Delta_0|^2 (S_\mathbf{k}^\perp)^2]^{1/2}.
$
We note that the quasiparticle energies reduce to
$
E_{s \mathbf{k} }  = \sqrt{\xi_\mathbf{k}^2 - |\Delta_0|^2} + s S_\mathbf{k}^\perp
$
when $\alpha_z = 0$,
$
E_{s \mathbf{k} }  = \sqrt{(\xi_\mathbf{k}+s S_\mathbf{k}^z)^2 - |\Delta_0|^2}
$
when $\alpha_x = \alpha_y = 0$, and
$
E_{s \mathbf{k} }  = \xi_{s \mathbf{k}} = \xi_\mathbf{k}+s S_\mathbf{k}
$
when $\Delta_0 \to 0$.

The mean-field self-consistency equations for $\Delta_0$ and $\mu$ are determined,
respectively, by setting $\partial \Omega_0/\partial |\Delta_0| = 0$ and 
$N_0 = - \partial \Omega_0/\partial \mu$, leading to either $\Delta_0 = 0$ 
or
\begin{align}
\label{eqn:gap}
\frac{1}{g} &= -\frac{1}{2}\sum_{s \mathbf{k} } 
\frac{\partial E_{s \mathbf{k} }}{\partial |\Delta_0|^2} \mathcal{X}_{s \mathbf{k} }, \\
\label{eqn:number}
N_0 &= - \frac{1}{2} \sum_{s \mathbf{k} } 
\left(1 +\frac{\partial E_{s \mathbf{k} }}{\partial \mu} \mathcal{X}_{s \mathbf{k} } \right).
\end{align}
Here, $N_0$ is the thermal average number of particles at the mean-field level,
$
\partial E_{s \mathbf{k} }/\partial |\Delta_0|
= -|\Delta_0| [1 + s (S_\mathbf{k}^\perp)^2/B_\mathbf{k}]/E_{s \mathbf{k} },
$
$
\partial E_{s \mathbf{k} }/\partial \mu
= - \xi_\mathbf{k} (1 + s S_\mathbf{k}^2/B_\mathbf{k})/E_{s \mathbf{k} },
$
and
$
\mathcal{X}_{s \mathbf{k} } = \coth[E_{s \mathbf{k} }/(2T)]
$
is a thermal factor.
Equations~(\ref{eqn:gap}) and (\ref{eqn:number}) follow from a Matsubara 
summation of the form
$
T\sum_{\ell} 1/(i\omega_\ell - x) = - n_B(x),
$
where
$
n_B(x) = 1/(e^{x/T} - 1)
$
is the Bose-Einstein distribution with
$
n_B(x) + n_B(-x) = -1
$
and
$
\coth[x/(2T)] = 1 + 2n_B(x).
$
It can be readily verified that all of these expressions recover the known 
counterparts in the absence of a SOC when 
$S_\mathbf{k} \to 0$~\cite{jeon02, koetsier09, qiang10}.
In addition, since a unidirectional SOC field in $\mathbf{k}$ space 
(e.g., $S_\mathbf{k} = |S_\mathbf{k}^i|$ for any $i \in \{x,y,z\}$) 
can be trivially gauged or integrated away from the self-consistency equations, 
it is identical to the $S_\mathbf{k} \to 0$ case up to an energy offset in $\mu$.

Following the standard prescription for the BCS-BEC crossover problem~\cite{strinati18}, 
we substitute the bare interaction strength $g$ between the $\uparrow$ and 
$\downarrow$ bosons with the associated $s$-wave scattering length $a_s$ 
in vacuum through the relation
$
1/g = -mV/(4\pi a_s) + \sum_\mathbf{k} 1/(2\epsilon_\mathbf{k}),
$
where $V$ is the volume. In addition, we define a length scale $k_0$ 
through an analogy with the number equation 
$
N_0 = k_0^3 V/(3\pi^2)
$
of a free Fermi gas at $T = 0$, along with the corresponding energy scale 
$\epsilon_0 = k_0^2/(2m)$.
Then, we solve Eqs.~(\ref{eqn:gap}) and (\ref{eqn:number}) for the saddle-point 
parameters $|\Delta_0|/\epsilon_0$ and $\mu/\epsilon_0$, and analyze their
stability as functions of $1/(k_0a_s)$, $T/\epsilon_0$, and $m\alpha/k_0$. 
The resultant phase diagrams are presented in Figs.~\ref{fig:as} and~\ref{fig:soc}, 
and they are constructed as follows.

\subsection{Critical pairing transition temperature}
Recalling that the thermodynamic stability of the paired state that is 
described by this mean-field theory requires a nonzero order parameter, 
we introduce an upper bound on $T$ that is based on the critical 
pairing transition temperature $T_p$, below which $\Delta_0 > 0$. 
Thus, by setting $\Delta_0 \to 0^+$ in Eqs.~(\ref{eqn:gap}) 
and~(\ref{eqn:number}), we find
$
1/g = \sum_{s \mathbf{k} } \mathcal{X}_{s \mathbf{k} }^p
[1 - s (S_\mathbf{k}^z)^2/(\xi_{s \mathbf{k} }S_\mathbf{k})]
/(4\xi_\mathbf{k})
$
for the $T_p$ equation, and 
$
N_0 = \sum_{s \mathbf{k} } n_B^p(\xi_{s \mathbf{k} })
$
for the number equation, where
$
\mathcal{X}_{s \mathbf{k} }^p = \coth[\xi_{s \mathbf{k} }/(2T_p)].
$
In addition, by requiring $\xi_{s\mathbf{k}} > 0$ in the entire $\mathbf{k}$ space 
for the normal state, we find that
$
|\mu| > m \alpha_m^2/2
$
with 
$
\alpha_m = \max\{ \alpha_x, \alpha_y, \alpha_x \}
$ 
corresponds to a lower bound on the stability of the mean-field 
$T_p$ equation. $\Delta_0 = 0$ below this bound, opening a window for 
the BEC of unpaired bosons.

\begin{figure}[htbp]
\includegraphics[scale=0.7]{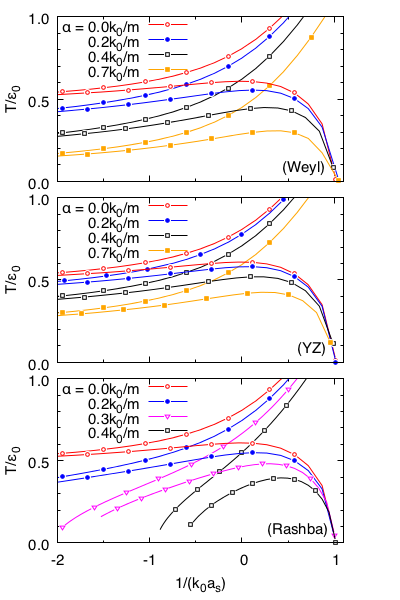}
\caption{
\label{fig:as} 
Finite temperature phase diagrams are constructed for the Weyl (upper 
panel), YZ (middle) and Rashba (lower) SOC fields. In each of these 
panels, we choose a set of SOC strengths $\alpha$, and vary the 
scattering length $a_s$. For a given $\alpha$, the BCS-type paired state 
is bounded by an upper and a lower curve, corresponding, respectively, 
to the critical transition temperature for the formation of pairs ($T_p$), 
and that of the gapless quasiparticle excitations ($T_g$). 
}
\end{figure}

In Fig.~\ref{fig:as}, our numerical calculations show that $T_p/\epsilon_0$ 
saturates to $0.5$ in the absence of a SOC in the weak-interaction limit 
when $g \to 0^+$ or $1/(k_0a_s) \to -\infty$.
This is in perfect agreement with our analytical derivations.
In addition, the mere effect of having a finite Weyl or YZ SOC on $T_p/\epsilon_0$ 
is seen to be the lowering of the saturation value in the strong-interaction 
limit when $1/(k_0a_s) \to +\infty$. 
It is pleasing to see that this $\alpha$-dependent effect gradually fades away 
with increasing $1/(k_0a_s)$, and $T_p/\epsilon_0$ eventually conforms to a 
SOC-independent growth in the opposite $1/(k_0a_s) \to +\infty$ limit. 
In sharp contrast, Fig.~\ref{fig:as} shows that having a finite Rashba SOC has 
a rather dramatic effect on $T_p/\epsilon_0$, and that the self-consistency 
equations do not yield a convergent solution once $1/(k_0a_s)$ is below 
an $\alpha$-dependent value. Thus, the pairing transition is discontinuous, 
and the precise location of the jumps in $T_g/\epsilon_0$ are determined 
by $|\mu| \to m\alpha^2/2$, signaling the violation of the $\xi_{s\mathbf{k}} > 0$ 
condition for the stability of the normal state.

\subsection{Critical gapless transition temperature}
In addition, for the dynamical stability of the fully paired state, it is necessary 
to restrict the self-consistent solutions to the parameter regime where the 
quasiparticle energies are real and positive in the entire $\mathbf{k}$ space. 
Note that $E_{s \mathbf{k} } \le 0$ may indicate a competition between a paired 
state and an unpaired one~\cite{koetsier09, qiang10}. Thus, by imposing 
the condition
$
E_{+,\mathbf{k}} E_{-,\mathbf{k}} = 0,
$
we find that the excitations are gapless in those $\mathbf{k}$-space regions 
satisfying
$
(\xi_\mathbf{k} + |\Delta_0| + S_\mathbf{k}^z) (\xi_\mathbf{k} - |\Delta_0| - S_\mathbf{k}^z )
= (S_\mathbf{k}^\perp)^2.
$
Then, by analyzing the gradient of $|\mu|$ in $\mathbf{k}$ space, we conclude 
that the gapless region is bounded by a minimum value of $|\mu|$ determined by
$
|\mu| = m \alpha_z^2/2 + |\Delta_0|
$
when
$
|\Delta_0| > m |\alpha_m^2 - \alpha_z^2|,
$
and by
$
|\mu| = m \alpha_m^2/2 + |\Delta_0|^2/(2m |\alpha_m^2 - \alpha_z^2|)
$
when
$
|\Delta_0| \le m |\alpha_m^2 - \alpha_z^2|.
$

\begin{figure}[htbp]
\includegraphics[scale=0.7]{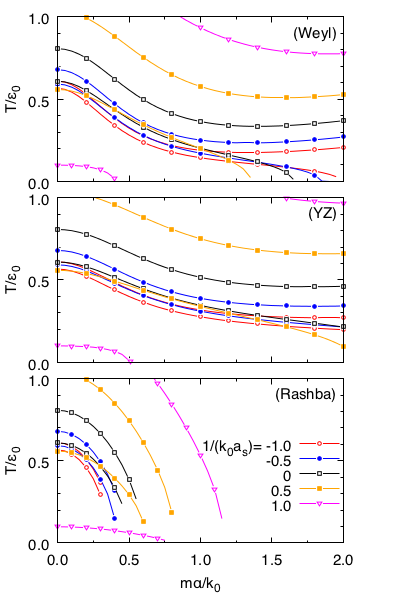}
\caption{
\label{fig:soc} 
Finite temperature phase diagrams are constructed for the Weyl (upper 
panel), YZ (middle) and Rashba (lower) SOC fields. In each of these 
panels, we choose a set of scattering lengths $a_s$, and vary the 
SOC strength $\alpha$. For a given $a_s$, the BCS-type paired state 
is bounded by an upper and a lower curve, corresponding, respectively, 
to the critical transition temperature for the formation of pairs ($T_p$), 
and that of the gapless quasiparticle excitations ($T_g$). 
}
\end{figure}

In order to identify this instability~\cite{koetsier09, qiang10}, 
here we introduce a lower bound on $T$
that is based on the critical gapless transition temperature $T_g$, above 
which $E_{s \mathbf{k} } > 0$. For instance, noting that the gapless 
transition condition reduces to $\mu = -|\Delta_0|$ in the absence of 
a SOC, one can analytically determine the precise location of $T_g \to 0^+$. 
We find
$
|\mu| = \pi^2/(16m a_s^2)
$
from the order parameter equation, and 
$
|\mu| = 2^{1/3} \epsilon_0
$
from the number equation, leading to a critical point
$
1/(k_0 a_s) = 2^{5/3}/\pi \approx 1.0106
$
that is in perfect agreement with our numerical calculations. 
Independently of the $\mathbf{k}$-space structure, Fig.~\ref{fig:as} shows that 
the precise location of the critical point is quite immune to the presence of 
a weak SOC, and that the ground ($T = 0$) state is stable for 
$1/(k_0 a_s) > 1.0106$ only. However, Fig.~\ref{fig:soc} shows that stronger 
SOCs eventually increase the stability of the ground state towards the lower 
$1/(k_0 a_s)$ regions as well. 
Going away from the critical point, one expects to find $T_g \to T_p$ from 
below as $\Delta_0 \to 0^+$ in the $1/(k_0a_s) \to -\infty$ limit, and this
turns out to be the case for a Weyl or a YZ SOC. In sharp contrast, 
Figs.~\ref{fig:as} and~\ref{fig:soc} again show peculiar jumps for the 
Rashba SOC, suggesting a discontinuous transition. 
We emphasize that these jumps are not numerical artifacts, and their origin 
can be traced back to the observation that self-consistency equations do not 
yield a convergent solution with $0 < |\Delta_0| < m\alpha^2$, and therefore, 
with $m\alpha^2/2 < |\mu| < m\alpha^2$. 

So far the mean-field analysis reveals that the $\mathbf{k}$-space structure 
of a SOC plays a crucial role in the boosted stability of the paired state at low $T$, 
but what makes a Rashba SOC distinct from a Weyl or a YZ SOC is yet to 
be uncovered. To address this question, next we analyze the Gaussian fluctuations 
of the order parameter around its mean-field, and gain more physical insight into 
the pairing problem.

\section{Gaussian fluctuations near $T_p$}
Going beyond the saddle-point approximation for pairing in the $\Delta_0 \to 0$ 
limit~\cite{melo93, koetsier09}, the Gaussian-fluctuation contribution 
$\Omega_G$ to the thermodynamic potential can be written as
$
\Omega_G =  \sum_q |\Lambda_q|^2/g 
- (T/4) \mathrm{Tr} \sum_{kq} G_0(k) \Sigma(q) G_0(k+q) \Sigma(-q).
$
Here, $q = (\mathbf{q}, i\nu_\ell)$ is a combined summation index 
with $\nu_\ell = 2\pi T \ell$ the bosonic Matsubara frequency for 
the pairs, $\Lambda_q$ is the spatial and temporal fluctuations of the 
order parameter around its saddle-point value, and
$
\Sigma(q) = \left[
\begin{matrix}
0 & \Lambda_q \sigma_x \\ 
\Lambda_{-q}^* \sigma_x & 0
\end{matrix}
\right].
$
A compact way to express 
$
\Omega_G = \sum_q \mathcal{L}_q^{-1} |\Lambda_q|^2
$
is through 
\begin{align}
\label{eqn:Lq}
\mathcal{L}_q^{-1} &= \frac{1}{g} - \frac{1}{8} \sum_{s s' \mathbf{k}} 
\frac{\mathcal{X}_{s, \mathbf{k}+\mathbf{q}/2} + \mathcal{X}_{s', -\mathbf{k}+\mathbf{q}/2}}
{\xi_{s, \mathbf{k}+\mathbf{q}/2} + \xi_{s', -\mathbf{k}+\mathbf{q}/2} + i\nu_\ell} 
C_{\mathbf{k}\mathbf{q}}^{ss'},
\end{align}
corresponding to the inverse of the fluctuation propagator associated 
with the pairs, where
$
\mathcal{X}_{s \mathbf{k} } = \coth[\xi_{s \mathbf{k} }/(2T)]
$
is a thermal factor, and
$
C_{\mathbf{k}\mathbf{q}}^{ss'} = 1 + ss'
(\mathbf{S}_{\mathbf{k}+\mathbf{q}/2} \cdot \mathbf{S}_{-\mathbf{k}+\mathbf{q}/2}
- 2S_{\mathbf{k}+\mathbf{q}/2}^z S_{-\mathbf{k}+\mathbf{q}/2}^z)/
(S_{\mathbf{k}+\mathbf{q}/2} S_{-\mathbf{k}+\mathbf{q}/2}).
$

By expanding the inverse propagator up to second order in the momentum
and first order in the frequency of the pairs, we find
$
\mathcal{L}_q^{-1} = a(T) + \frac{1}{2} \sum_{ij} c_{ij} q_i q_j - d \omega + \dots.
$
This is our Ginzburg-Landau theory in disguise~\cite{melo93}, whose zeroth-order term 
$
a(T) = \mathcal{L}_0^{-1}
$
determines the transition temperature, the second-order kinetic coefficient
$
c_{ij} = \lim_{q \to (\mathbf{0},0)} \partial^2 \mathcal{L}_q^{-1}/(\partial q_i \partial q_j)
$
is related to the effective mass tensor $\mathbf{m_p}$ of the pairs, 
and the first-order coefficient
$
d \omega \stackrel{\omega \to 0}{=} \mathcal{L}_0^{-1} - \lim_{q \to (\mathbf{0},-\omega+i0^+)} \mathcal{L}_q^{-1}
$
characterizes the dynamical stability or lifetime of the pairs.
Note that the zeroth order term can also be written as
$
a(T) = \lim_{\Delta_0 \to 0} \partial \Omega_0/\partial |\Delta_0|^2,
$
showing that the Thouless condition
$
a(T_p) = 0
$
reproduces the equation for $T_p$.
Similarly, going beyond the Gaussian fluctuations, the zeroth-order coefficient
of the fourth-order fluctuations in $\Lambda(q)$ can be approximated as
$
b = \lim_{\Delta_0 \to 0} \partial^2 \Omega_0/\partial (|\Delta_0|^2)^2.
$
This higher-order coefficient controls the interaction strength $g_p = b/d^2$ 
between pairs, where
$
b =  \lim_{\Delta_0 \to 0} \sum_{s \mathbf{k} } \{
[\partial^2 E_{s \mathbf{k} } /\partial (|\Delta_0|^2)^2] \mathcal{X}_{s \mathbf{k} }/2
- (\partial E_{s \mathbf{k} }/\partial |\Delta_0|^2)^2 \mathcal{Y}_{s \mathbf{k} }/(4T)
\}
$
with 
$
\mathcal{Y}_{s \mathbf{k}} = \cosech^2[\xi_{s \mathbf{k} }/(2T)]
$
an additional thermal factor. 

The presence of thermal factors makes the coefficients of the Ginzburg-Landau 
theory rather cumbersome for the many-body problem, and this holds true even
in the absence of a SOC. However, it is possible to circumvent around this 
complication in the two-body limit and make further analytical progress.
For instance, for the two-body binding problem in vacuum where $\mu < 0$ 
with $|\mu| = \epsilon_b/2 \gg T_p \to 0$, by setting the thermal factors to 
unity (i.e., $\mathcal{X}_{s \mathbf{k} } \to 1$ for every $s \mathbf{k}$ assuming 
$\xi_{s\mathbf{k}} > 0$) and introducing $\epsilon_b = i\nu_\ell - 2\mu$ as 
the binding energy at $T = 0$, we find
$
\mathcal{L}_\textrm{tb}^{-1} = 1/g - (1/4) \sum_{s s' \mathbf{k}} 
C_{\mathbf{k}\mathbf{q}}^{ss'}/ (\epsilon_{s, \mathbf{k}+\mathbf{q}/2} + \epsilon_{s', -\mathbf{k}+\mathbf{q}/2} + \epsilon_b). 
$
This is the inverse propagator for the two-body bound states, and it can be 
used to extract both the binding energy $\epsilon_b$ and the effective mass 
tensor $\mathbf{m_p}$ of the pairs as follows.

\subsection{Binding energy of the two-body bound state}
By applying the Thouless condition for the two-body problem $a_{\textrm{tb}}(0) = 0$,
we find
\begin{align}
\label{eqn:eb}
\frac{1}{g} = \sum_\mathbf{k} \frac{(2\epsilon_\mathbf{k}+\epsilon_b)^2-4(S_\mathbf{k}^\perp)^2}
{(2\epsilon_\mathbf{k}+\epsilon_b)[(2\epsilon_\mathbf{k}+\epsilon_b)^2-4S_\mathbf{k}^2]},
\end{align}
which is analytically tractable in various limits. For instance, when 
$S_\mathbf{k}^\perp = 0$, Eq.~(\ref{eqn:eb}) suggests that a bound state with 
$\epsilon_b = 1/(m a_s^2)$ exists only for $a_s > 0$. This result is not surprising 
given that a unidirectional SOC field in $\mathbf{k}$ space (e.g., the remaining 
SOC component $S_\mathbf{k}^z$) can simply be gauged away even from 
our many-body mean-field. More intriguingly, for a Rashba SOC, Eq.~(\ref{eqn:eb}) 
again suggests that a bound state with $\epsilon_b = 1/(m a_s^2)$ exists only for 
$a_s > 0$, and that the presence of a Rashba SOC does not have any effect on the 
binding energy. This null result is quite surprising given its non-trivial fermionic 
counterpart, where a bound state with $\epsilon_b \ne 0$ is known to exist for 
any $a_s$ no matter how weak $g$ is as long as $g \ne 0$~\cite{iskin11}.
However, we note that the intra-component pairing of bosons with Rashba 
SOC is similar to the fermion problem~\cite{li14}.

On the contrary, for a Weyl SOC, Eq.~(\ref{eqn:eb}) suggests that a bound state 
with $\epsilon_b \ne 0$ exists for any $a_s$, according to
$
3/(m\alpha a_s) = 2\sqrt{\epsilon_b/(m\alpha^2)} + \sqrt{\epsilon_b/(m\alpha^2)-1}
-1/\sqrt{\epsilon_b/(m\alpha^2)-1}.
$
This leads to
$
\epsilon_b = m\alpha^2 [1 + (m \alpha a_s/3)^2]
$
in the weak-binding limit when $\epsilon_b \to m\alpha^2$ or 
$1/(m \alpha a_s) \to -\infty$,
$
\epsilon_b = 2m\alpha^2/\sqrt{3} = 1.1547 m\alpha^2
$
at unitarity when $|a_s| \to \infty$, and
$
\epsilon_b = m\alpha^2 + 1/(m a_s^2)
$
in the strong-binding limit when $\epsilon_b \gg m \alpha^2$ or 
$1/(m \alpha a_s) \to +\infty$. Thus, our formalism recovers the exact
solution of the two-body problem known for a Weyl SOC~\cite{luo17}.
Similarly, for a YZ SOC, Eq.~(\ref{eqn:eb}) again suggests that a bound state 
with $\epsilon_b \ne 0$ exists for any $a_s$, according to 
$
2/(m\alpha a_s) = 2\sqrt{\epsilon_b/(m\alpha^2)} 
- \asinh[1/\sqrt{\epsilon_b/(m\alpha^2)-1}].
$
This leads to
$
\epsilon_b = m\alpha^2 [1 + 4 e^{4/(m \alpha a_s)-4}]
$
in the $1/(m \alpha a_s) \to -\infty$ limit,
$
\epsilon_b = 1.06640 m\alpha^2
$
at unitarity, and
$
\epsilon_b = m\alpha^2 + 1/(m a_s^2)
$
in the $1/(m \alpha a_s) \to +\infty$ limit. 

Given these analytical results for the two-body problem, we conclude that it is 
the coupling between $S_\mathbf{k}^z$ and the other components 
($S_\mathbf{k}^x$ and/or $S_\mathbf{k}^y$) that gives rise to a two-body
bound state with $\epsilon_b \ne 0$ for any $a_s < 0$ as long as $g \ne 0$.
This conclusion clearly sheds some light on the boosted stability of the many-body 
problem in general, and particularly on Figs.~\ref{fig:as} and~\ref{fig:soc}.

\subsection{Effective mass of the two-body bound state}
It turns out that the elements $(\mathbf{m_p^{-1}})^{ij} = c_{ij}/d$ of the 
inverse-effective-mass tensor are also analytically tractable for the two-body problem. 
For instance, in the weak-binding limit when $1/(m\alpha a_s) \to -\infty$, 
we note that $S_\mathbf{k}^z \ne 0$ for the existence of a two-body bound 
state to begin with, and find
$
c_{ij} \to \sum_{s \mathbf{k} } [\partial^2 \xi_{s \mathbf{k} }/(\partial k_i \partial k_j)] 
(S_\mathbf{k}^z)^2 / (16\xi_{s \mathbf{k} }^2 S_\mathbf{k}^2)
$
and
$
d \to \sum_{s \mathbf{k} } (S_\mathbf{k}^z)^2/(8 \xi_{s \mathbf{k} }^2 S_\mathbf{k}^2).
$
For a Weyl SOC, setting $\epsilon_b \to m\alpha^2$ in the $1/(m \alpha a_s) \to -\infty$ 
limit, we find
$
d = mV\sqrt{m}|\mu| / [12\pi(2|\mu|-m\alpha^2)^{3/2}],
$
leading to a diagonal mass tensor with anisotropic elements 
$m_p^{xx} = m_p^{yy} = 10m$ and $m_p^{zz} = 10m/3$. 
This result is in sharp contrast with its fermionic counterpart, where 
$m_p^{xx} = m_p^{yy} = m_p^{zz} = 6m$ is known to be isotropic in the entire 
space~\cite{iskin11}. Similarly, for a YZ SOC, setting again $\epsilon_b \to m\alpha^2$ 
in the $1/(m \alpha a_s) \to -\infty$ limit, we find
$
d = mV\sqrt{2m|\mu|} / [16\pi(2|\mu|-m\alpha^2)],
$
leading again to a diagonal mass tensor with anisotropic elements 
$m_p^{xx} = 2m$, $m_p^{yy} = 8m$ and $m_p^{zz} = 8m/3$. 
This result is again in sharp contrast with its fermionic counterpart, where 
$m_p^{xx} = 2m$ but $m_p^{yy} = m_p^{zz} = 4m$ is known to be isotropic 
in $yz$ plane~\cite{iskin11}. We note that a Rashba SOC gives rise to 
an anisotropic mass tensor for the pairs, when the pairing is due to 
the intra-component attraction~\cite{li14}.

As a result of this two-body analysis, we conclude that the coupling between
$S_\mathbf{k}^z$ and the other components ($S_\mathbf{k}^x$ and/or $S_\mathbf{k}^y$) 
gives rise to an anisotropic $\mathbf{m_p}$ in general for the many-body 
bound states~\cite{notemB}. One can show that the SOC-induced anisotropy 
gradually disappears with increasing $\epsilon_b$, in such a way that
$
m_p^{xx} = m_p^{yy} = m_p^{zz} = 2m
$
is eventually isotropic in space in the strong-binding limit when 
$1/(m\alpha a_s) \to +\infty$. Note that the effective mass tensor of pairs plays 
a direct role in their finite $T$ phase diagrams. For instance, the critical BEC 
temperature $T_c$ of non-interacting pairs is determined by their number equation
$
N_p = \sum_{\mathbf{k}}n_B^c(\epsilon_{p\mathbf{k}}),
$
and plugging the anisotropic dispersion $\epsilon_{p\mathbf{k}}$ for the free pairs, 
we approximate
$
T_c = 0.218  \epsilon_0 \times 2m/(m_p^{xx}m_p^{yy}m_p^{zz})^{1/3}
$
in the $g_p \to 0$ limit~\cite{iskin11}.
Thus, while $T_c/\epsilon_0$ saturates to $0.0629$ for a Weyl SOC and 
to $0.1248$ for a YZ SOC in the $1/(m \alpha a_s) \to -\infty$ limit, it reduces 
to the usual result $0.218$ in the $1/(m \alpha a_s) \to +\infty$ limit.

\section{Summary}
Here we analyzed the properties of BCS-type paired state in a 
spin-$\frac{1}{2}$ Bose gas with arbitrary SOC. Relying on the 
mean-field and Ginzburg-Landau theories for the paired state, we showed 
how the $\mathbf{k}$-space structure of a SOC field manifests in the many-body 
and two-body problems, boosting the stability of the paired state as a 
function of its strength. For this purpose, we calculated the critical transition 
temperature for the formation of pairs, and that of the gapless quasiparticle 
excitations for a broad range of interaction and SOC strengths. 
It turns out that while the SOC has a minor role in the strong-interaction 
limit where the ground state at $T = 0$ is a paired BEC, increasing its 
strength in the weak-interaction limit may allow for the creation of a 
paired state at much lower temperatures. We also provided a full account 
of the two-body problem including its non-vanishing binding energy for 
arbitrarily weak interactions and the anisotropic effective mass tensor.

\begin{acknowledgments}
This work is supported by the funding from T{\"U}B{\.I}TAK Grant No. 1001-118F359.
\end{acknowledgments}


\begin{thebibliography}{99}
\bibitem{mueff}
The Hartree terms are taken into account through an energy shift in the chemical 
potential of the $\sigma$ component, i.e.,
$
\mu_\sigma \to \mu_\sigma - (U_{\uparrow \downarrow}+U_{\downarrow \uparrow})N_{-\sigma}/2
+ 2U_{\sigma \sigma} N_\sigma,
$
where $N_\sigma$ is the number of $\sigma$ particles. In this work, we simply have
$
\mu \to \mu - V_H
$
with $V_H = (U-g/2)N_0$ for the many-body problem, assuming 
$N_\uparrow = N_\downarrow = N_0/2$ and 
$U_{\uparrow \uparrow} = U_{\downarrow \downarrow} = U > 0$. 
For instance, one of the conditions for the mechanical stability is a positive isothermal 
compressibility, and our mean-field theory satisfies $\partial \mu/\partial N_0 > 0$ 
for $U$ that is sufficiently larger than $g/2$.
\bibitem{muB}
The role played by the Hartree shift $V_H$ can also be seen as follows. 
Due to its analytical simplicity, let us set the SOC to 0, and consider the 
strong-coupling limit of Eqs.~(\ref{eqn:gap}) and~(\ref{eqn:number}). 
Since $|\mu| \to \epsilon_b/2 \gg |\Delta_0|$, we find that 
$
N \approx \sum_\mathbf{k} |\Delta_0|^2/(2\xi_\mathbf{k}^2)
$
leading to
$
|\Delta_0|^2 = [16/(3\pi)] \sqrt{\epsilon_0^3 |\mu|},
$
and 
$
1/g \approx \sum_\mathbf{k} [1/(2\xi_\mathbf{k}) + |\Delta_0|^2/(4\xi_\mathbf{k}^3)]
$
leading to
$
1/(k_0 a_s) = \sqrt{|\mu|/\epsilon_0} - |\Delta_0|^2/(16 \sqrt{\epsilon_0 |\mu|^3}).
$
When we take $V_H$ into account and identify
$
\mu_p = 2\mu + \epsilon_b = 2V_H - 4\epsilon_0 k_0a_s/(3\pi)
$
as the chemical potential of the pairs, and then match it with 
$g_p N_p$ of weakly-interacting pairs where $g_p = 4\pi a_p/(m_p V)$, 
$N_p = N_0/2$ and $m_p = 2m$, we find that the effective pair-pair 
scattering length is given by $a_p = 2mV (U-g/2)/\pi - 2a_s$. Thus, for 
$U$ that is sufficiently larger than $g/2$, we find that $a_p$ is positive and 
the system is not prone to collapse. In the absence of a Hartree shift, 
we note that $a_p = - 2a_s < 0$ is in agreement with that of the 
spinless case~\cite{koetsier09}, and is in sharp contrast with its fermionic 
counterpart $a_p = 2a_s > 0$~\cite{melo93}.
\bibitem{notemB} For a two-dimensional spin-$\frac{1}{2}$ Bose gas 
with a two-dimensional SOC, we find that the mass tensor is independent of 
the symmetry of the SOC fields in the weak-binding limit, i.e., 
$m_p^{ij} = 2m \delta_{ij}$ for the relevant dimensions for an XY, XZ or a YZ SOC.
\end{thebibliography}
\end{document}